\documentclass{ws-procs961x669}            
\begin{document}
\title{Gaining confidence on general relativity\\
with cosmic polarization rotation}

\author{Sperello di Serego Alighieri}

\address{INAF - Osservatorio Astrofisico di Arcetri\\
Largo E. Fermi 5, 50125 Firenze, Italy\\
E-mail: sperello@arcetri.astro.it\\
www.arcetri.astro.it/$\sim$sperello}

\begin{abstract}
The theory of general relativity, for which we celebrate the centennial at this Symposium, is based on the Einstein equivalence principle. This principle could be violated through a pseudoscalar-photon interaction, which would also produce a rotation of the polarization angle for radiation traveling over very long distances. Therefore, if we could show that this cosmic polarization rotation does not exist, our confindence in general relativity would be greatly increased. We review here the astrophysical searches for cosmic polarization rotation, which have been made in the past 26 years using the polarization of radio galaxies and of the cosmic microwave background. So far no rotation has been detected within about 1 degree. We discuss current problems and future prospects for cosmic polarization rotation measurements.
\end{abstract}


\bodymatter

\section{The importance of cosmic polarization rotation}\label{aba:sec1}

The possibility that space-time is curved by a gravitational field, as postulated by the theory of general relativity (GR), rests on the extension of the equivalence between an accelerated frame and a gravitational field from experiments involving just moving bodies (the weak or Galilean equivalence principle, WEP\cite{Gal38}) to those concerning all non-gravitational forces, including electromagnetic fields and photons (the Einstein equivalence principle, EEP\cite{Ein07}). In fact, the trajectory of a photon in an accelerated frame is curved because of the finite speed of light, and the same should equivalently occur in a gravitational field. However, this extension from the WEP to the EEP would not be justified if there were a pseudoscalar field $\phi$ coupling to electromagnetism and leading to a violation of the EEP, while obeying the WEP.\cite{Nio73, Nio77} Such pseudoscalar field, if it existed, would also produce a rotation of the polarization angle (PA) for radiation traveling over large distances across the universe,\cite{Car98} the so called cosmic polarization rotation (CPR). If we could show that there is no CPR, the EEP would be tested with the same high accuracy of the WEP, greatly increasing our confidence in the EEP and consequently in GR. In fact CPR could also be induced by violations of other fundamental physical principles (see ref.~\refcite{Nio10} for a recent review), connected with the breaking of symmetry and parity: clearly, if there were any CPR, i.e. if the CPR angle $\alpha$ were not zero, it should be either positive for a counter-clockwise rotation, or negative for a clockwise rotation,\footnote{We adopt the convention for PA enforced by the International Astronomical Union (IAU): it increases counter-clockwise facing the source, from North through East.\cite{iau74}} leading to asymmetry.

Searches for CPR are important also to make the best use of the information about the universe which photons carry to us. Almost all the information we have about the universe outside the Solar System is carried to us by photons; a few cosmic rays, several elusive neutrinos, and the just discovered gravitational waves are the only exceptions. Photons carry information about their direction, their energy (or wavelength/frequency), and their polarization. The latter consists essentially in the position angle (PA) of the polarization ellipse, i.e. photons carry throughout the universe an important geometrical information, although our eyes cannot see it. In order to make the best use of this information, it is important to know if and how it is changed while photons travel to us. We know that the direction of photons can be modified by a strong gravitational field, and that their energy is modified by the expansion of the universe. Is the polarization PA also modified while photons travel large distances in vacuum?\footnote{We know that the polarization PA is modified while photons travel in a plasma with a magnetic field, the so called Faraday rotation, which is proportional to the wavelength squared. However we deal here only with possible modifications in vacuum.} The searches for CPR deal with this important question.

In the following I will briefly review the tests on CPR carried out with different methods in the past 26 years, and discuss current problems and future prospects for these tests. A more extensive recent review of CPR tests can be found in ref.~\refcite{diS15}.

\section{Searches for cosmic polarization rotation}

CPR tests are simple in principle: they require a distant source of polarized radiation for which the polarization orientation $PA_{em}$ at the emission or at the last interaction can be established. By measuring the observed orientation $PA_{obs}$, the CPR angle can be calculated:
$$\alpha  = PA_{obs} - PA_{em}. $$
The problem is the estimate of $PA_{em}$, since it is not generally possible to know the geometry of the emission process. Fortunately this problem can be solved when the last interaction was due to scattering and when the scattering geometry can be figured out. In this case we can use the fact that scattered radiation is polarized perpendicularly to the plane containing the incident and scattered rays. This simple physical law has been applied to CPR tests, using both the ultraviolet (UV) radiation of radio galaxies (RG) and the tiny disuniformities of the cosmic microwave background (CMB). The first CPR tests 26 years ago used instead a stastical analysis of the radio polarization in RG.\cite{Car90} The most accurate CPR tests obtained with the various methods are summarized in Fig.~1, which is derived from the data of ref.~\refcite{diS15}.

\begin{figure}[h]
\begin{center}
\includegraphics[width=12cm]{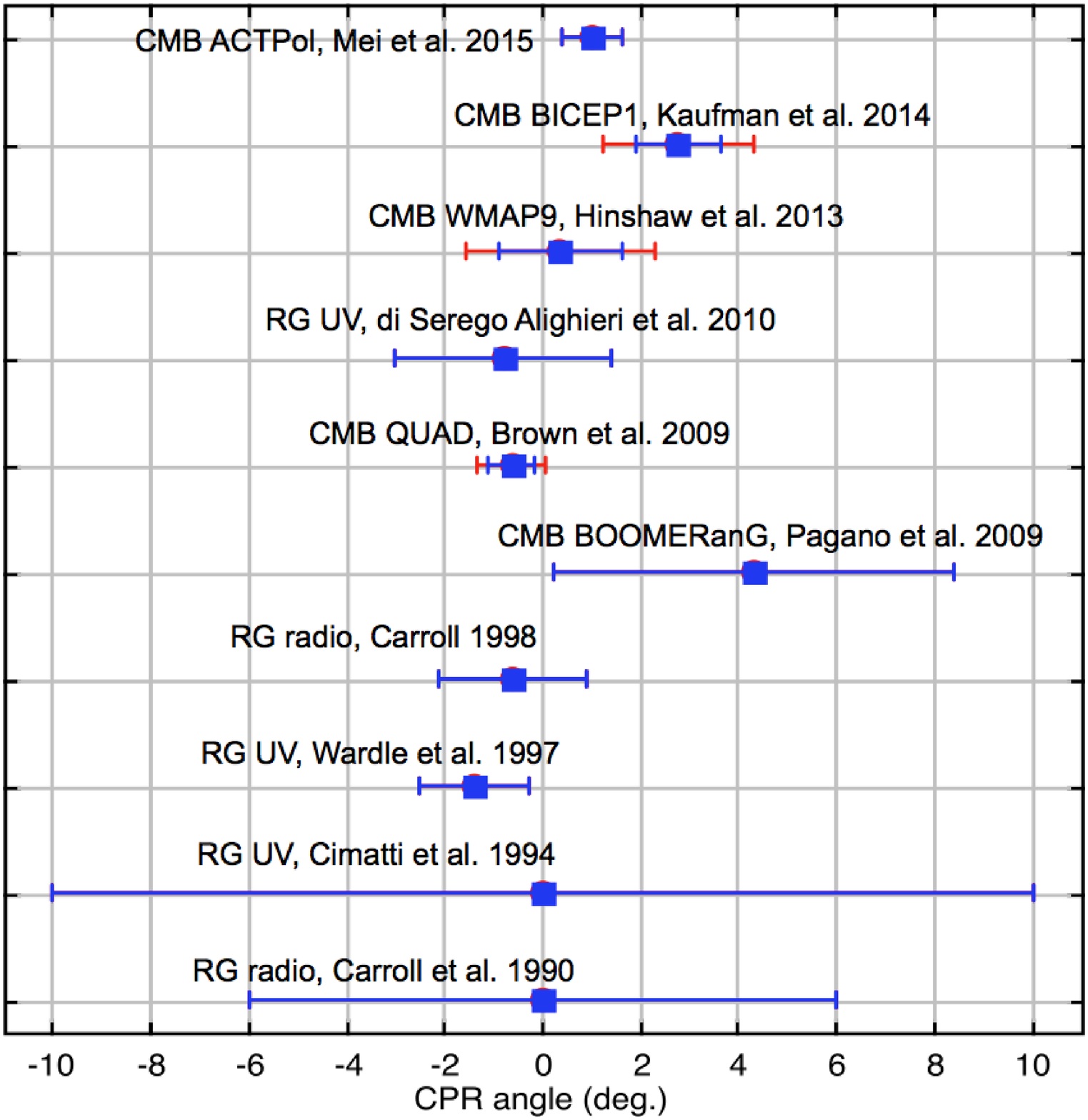}
\end{center}
\caption{CPR angle measurements by the various experiments, displayed in chronological order. Blue error bars are for the statistical error, while red ones include also the systematic one, if present. A systematic error should be added to the ATCPol measurement, equal to the unknown difference of the Crab Nebula polarization PA between 146 GHz and 89 GHz.\cite{Mei15}}
\label{aba:fig1}
\end{figure}

Since this review\cite{diS15} some new measurements connected with CPR have become available. The POLARBEAR collaboration\cite{Ade14b} reports about a difference of $1.08^{\circ}$ in the instrument polarization angle obtained minimizing the EB spectrum and that obtained from the Crab Nebula using the PA measurement of ref.~\refcite{Aum10}. This corresponds to a measurement of CPR, performed with the effect of a rotation on the EB spectrum and using the Crab Nebula for the PA calibration, and giving a CPR angle $\alpha = 1.08^{\circ} \pm 0.2^{\circ} (stat.) \pm 0.5^{\circ} (syst.)$. The BICEP2 collaboration\cite{Ade15} finds that TB and EB spectra at high monopoles are consistent with a coherent rotation of the polarization angle by $1.1^{\circ}$ with a random error of $0.2^{\circ}$. However this is not necessarily a measurement of the CPR angle $\alpha$, since this TB and EB minimization procedure\cite{Kea13} cannot disentangle $\alpha$ from a PA systematic error due to calibration uncertainties (see the next section). Recently the first results about CPR from the Planck satellite have become available: a CPR angle $\alpha = 0.0^{\circ} \pm 1.3^{\circ} (stat.) \pm 1^{\circ} (syst.)$ has been obtained from an analysis of just TE and EE spectra.\cite{Gru15} However this is not the final word on CPR from Planck, since, when also the BB, TB, and EB spectra will be included in the analysis, it is foreseen that the statistical uncertainty on the CPR angle should decrease to about $0.03^{\circ}$,\cite{Gru15} and hopefully it will also be possible to decrease the systematic errors with an improved PA calibration.

The CPR tests considered above assume that the rotation angle $\alpha$ is uniform across the sky. However it is also possible that $\alpha$ depends on the direction one is looking at\cite{Nio05} and that one can therefore measure its fluctuations $\langle \delta \alpha ^2 \rangle ^{1/2}$. The CPR data obtained on 8 RG\cite{diS10} have been used to constrain $\langle \delta \alpha ^2 \rangle ^{1/2}\leq 3.7^{\circ}$.\cite{Kam10} A better constraint $\langle \delta \alpha ^2 \rangle ^{1/2}\lesssim 1^{\circ}$ has been derived from the WMAP7 data.\cite{Glu12} B-mode CMB polarization data have been used to set a constraint $\langle \delta \alpha ^2 \rangle ^{1/2}\leq 1.56^{\circ}$.\cite{diS14}

Attempts have been made to derive constraints on CPR by combining data from various CMB experiments. For example, a CPR angle of 2.12 $\pm 1.14^{\circ}$ has been obtained\cite{Lio15} combining the data of BOOMERanG,\cite{Pag09} WMAP9,\cite{Hin13} and BICEP1.\cite{Kau14} However we find it problematic to combine data taken at different frequencies, in different areas of the sky, and particularly in the presence of large and different systematic errors due to uncertainties in the PA calibration.

In summary, the results so far are consistent with a null CPR with upper limits of the order of one degree.

\section{Current problems in searches for cosmic polarization rotation}

Searches for CPR using the UV polarization of RG have reached the limits allowed by current instrumentation, for the lack of suitable RG, for which the test can be performed and which are bright enough so that their polarization can be measured with the available instruments.

The most accurate results are now obtained with the CMB polarization, averaging over large sky areas, i.e. assuming uniform CPR over these areas. A current problem with CPR searches using the CMB is the calibration of the polarization PA for the lack of sources with precisely known PA at CMB frequencies. This introduces a systematic error, which is similar to the statistical measurement error, of the order of 1 degree (see the red error bars in Fig.~1). Recently the polarization PA of the Crab Nebula (Tau A) for a beam of 5' and 10' has been measured with an accuracy of $0.2^{\circ}$ at 89.2 GHz.\cite{Aum10} However most CMB polarization measurements are made at higher frequencies (100--150 GHz) and the Crab Nebula is not visible from the South Pole, the site of several CMB experiments. In order to overcome the PA calibration problem, some CMB polarization experiments have used a TB and EB nulling procedure.\cite{Kea13} However this procedure would eliminate together the PA systematic error and any CPR angle $\alpha$, so it cannot be used for CPR tests.

A second problem with CMB tests of CPR is that unfortunately the CMB polarimetrists have adopted the convention that the polarization PA increases clockwise (looking at the source), which is opposite to the standard convention adopted by all other polarimetrists for centuries and enforced by the IAU (PA increases counter-clockwise).\cite{iau74} This corresponds to a change of the sign of the U Stokes parameter and is obviously producing problems when comparing measurements with different methods, like for CPR tests, also because the “CMB convention” has not been well documented in the CMB polarization papers. 
The \lq\lq CMB convention" has been unfortunately adopted following a software for pixelization on a sphere.\cite{Gor05} However a pixelization software using the IAU convention exists\cite{Wal15} and changing the sign of the U Stokes parameter does not require any sophisticated software and can be done easily.\cite{Ade16} Recently the IAU has issued a recommendation that all astronomers, including those working on the CMB, follow the IAU convention for the PA.\cite{iau15} It would clearly be desirable that we all use the same convention and avoid parochialism in Science. 

\section{Summary and future prospects}

All results are so far consistent with a null CPR. The CPR test methods have reached so an accuracy of the order of $1^{\circ}$ and upper limits to any rotation of the same order. 

The different methods are complementary in many ways. They cover different wavelength ranges and the methods at shorter wavelength have an advantage, if CPR effects grow with photon energy, as foreseen in some cases\cite{Kos01, Kos02} . They also reach different distances, and the CMB method obviously reaches furthest. However the relative difference in light travel time between z = 3 and z = 1100 is only 16\%.

Improvements can be expected by better targeted high resolution radio polarization measurements of RGs and quasars, by more accurate UV polarization measurements of RGs with the coming generation of giant optical telescopes,\cite{deZ14, San13, Ber14} and by future CMB polarization measurements such as those from BICEP3.\cite{Ahm14}. Also the final results by the Planck satellite\cite{Ade14a} are foreseen to bring improvements for CPR measurements because of the small expected statistical error (see Section 2). In any case, in order to exploit its great accuracy, Planck will have to reduce accordingly also the systematic error in the calibration of the polarization angle, which at the moment is of the order of $1^{\circ}$ for the CMB polarization experiments. The best prospects to achieve this improvement are likely to be more precise measurements of the polarization angle of celestial sources at CMB frequencies with ATCA\cite{Mas13} and with ALMA,\cite{Tes13} and a calibration source on a satellite.\cite{Kau16}

\bibliographystyle{ws-procs961x669}
\bibliography{ws-pro-sample}



\end{document}